%% file: NIAGARA-pearc19.tex
\documentclass[sigconf]{acmart}

\def\BibTeX{{\rm B\kern-.05em{\sc i\kern-.025em b}\kern-.08emT\kern-.1667em\lower.7ex\hbox{E}\kern-.125emX}}
    
\input{ACMcopyright}

\usepackage{colortbl}
\usepackage{verbatim}
\usepackage{listings}
\usepackage{booktabs}
\usepackage{hyperref}
\usepackage{enumitem}

\begin{document}

\title[Deployment of the Niagara Supercomputer for Large Parallel Workloads]{Deploying a Top-100 Supercomputer for Large Parallel Workloads: the Niagara Supercomputer}

\input{authors1}

\renewcommand{\shortauthors}{M. Ponce et al.}

\begin{abstract}
Niagara is currently the fastest supercomputer accessible to academics in Canada.
It was deployed at the beginning of 2018 and has been serving the research community ever since.
This homogeneous 60,000-core cluster, owned by the University of Toronto and operated by SciNet,
was intended to enable large parallel jobs and has a measured performance
of 3.02 petaflops, debuting at \#53 in the June 2018 TOP500 list. 
It was designed to optimize throughput of a range of scientific codes running at scale,
energy efficiency, and network and storage performance and capacity. 
It replaced two systems that SciNet operated for over 8 years, the Tightly Coupled
System (TCS) and the General Purpose Cluster (GPC) \cite{scinet-gpc}.
In this paper we describe the transition process from these two systems, the procurement and
deployment processes, as well as the unique features that make Niagara
a one-of-a-kind machine in Canada.
\end{abstract}




\begin{teaserfigure}
  \centering
  \includegraphics[height=0.185\textwidth]{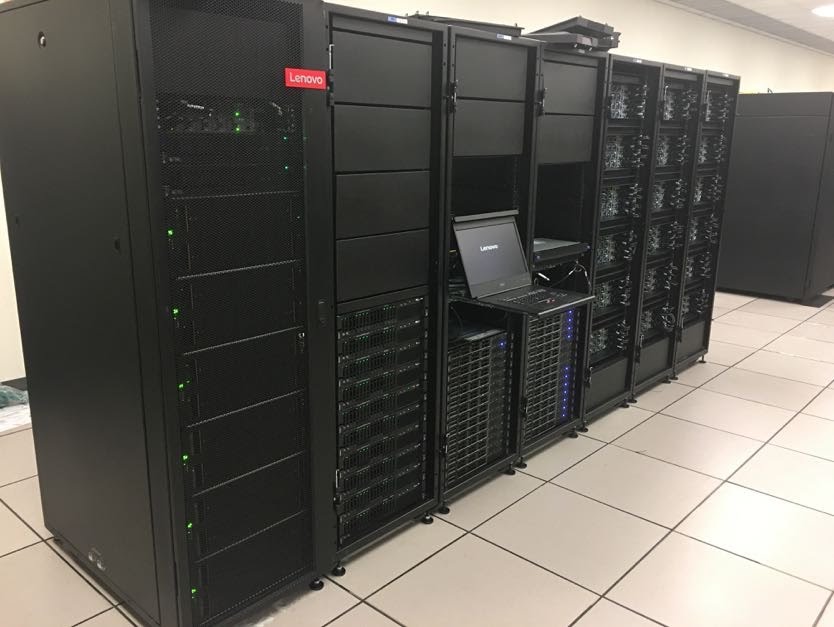}
  \includegraphics[height=0.185\textwidth]{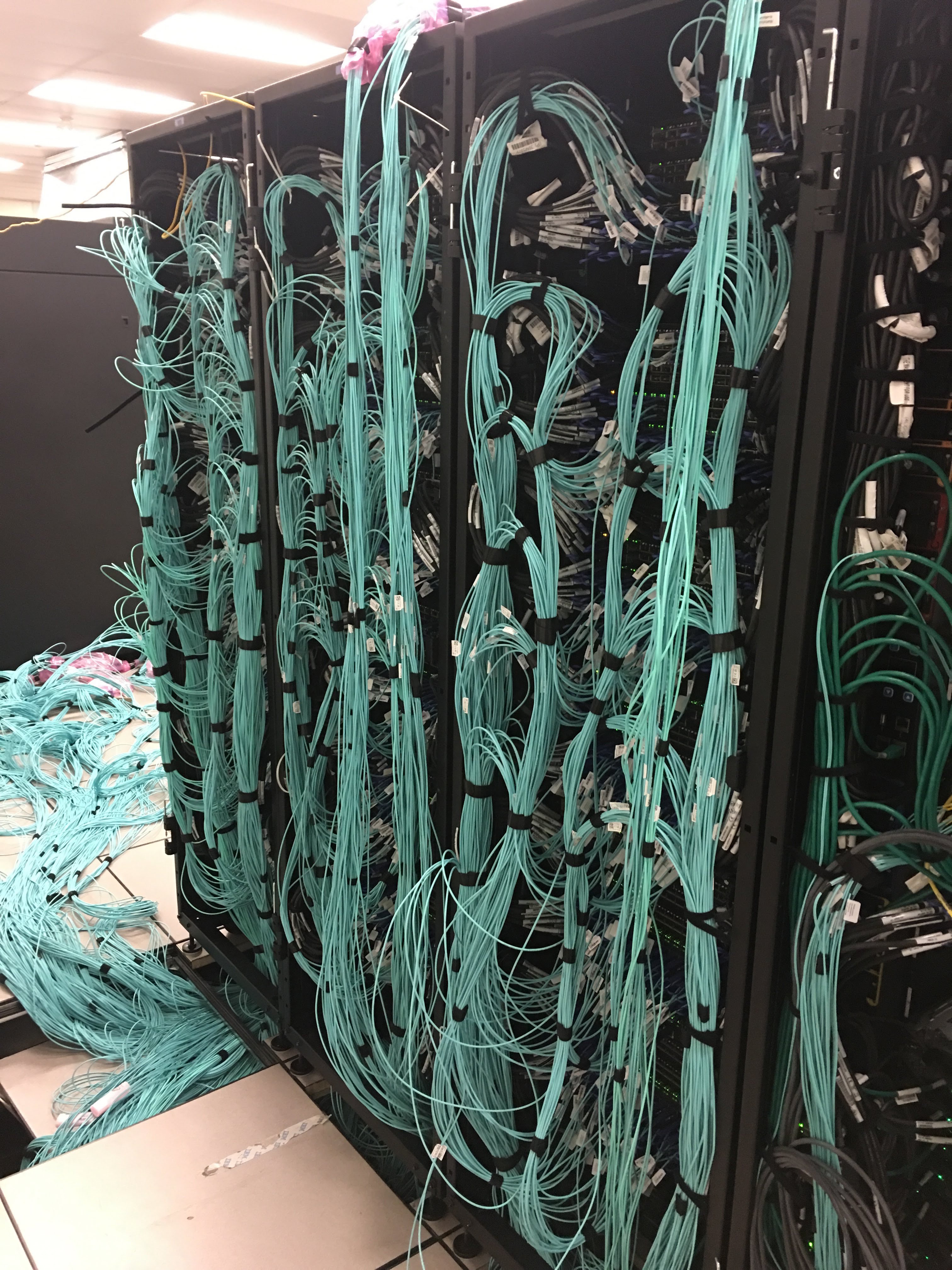}
  \includegraphics[height=0.185\textwidth]{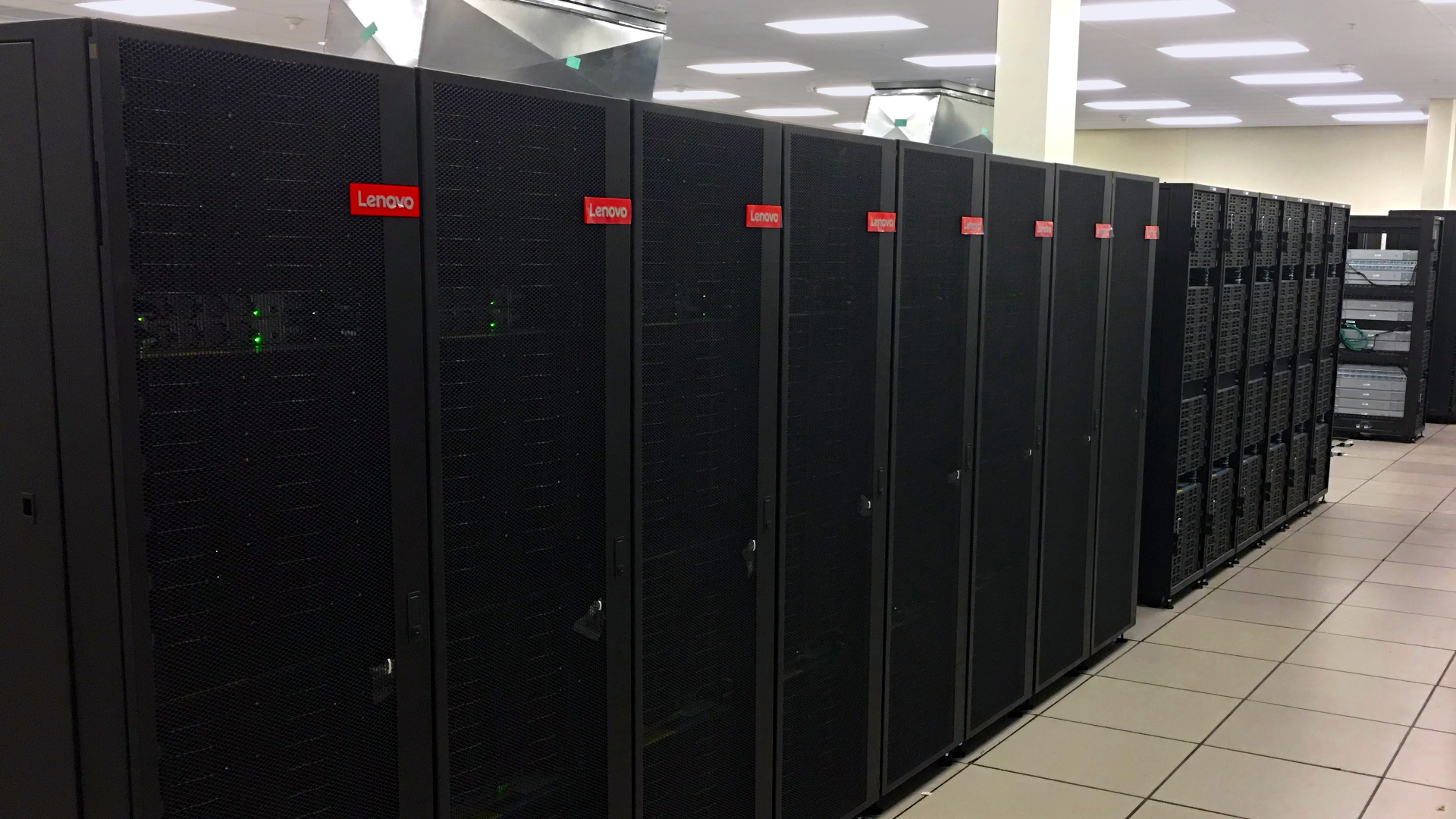}
  \caption{SciNet's Niagara supercomputer at the University of Toronto.
	From left to right: management nodes, Dragonfly interconnect, racks with compute nodes.}
  \Description{Niagara is the largest supercomputer available for researchers in Canada}
  \label{fig:teaser}
\end{teaserfigure}

\maketitle

\section{Introduction}
In this paper we describe the design, procurement, deployment,
installation, testing and setup for the Niagara supercomputer.
We focus on the ``best practices'' followed and developed for this specific
state-of-the-art homogeneous cluster, which could be of benefit to others
about to undertake a similar path.
Moreover, we will discuss the early science cases used to test,
validate, and fine-tune the system by attempting ``heroic'' full-system runs \cite{heroicRun},
which have already produced scientific results \cite{2018arXiv181013416W,HerwigRD},
and one of which received the
``2018 HPCwire Editor's Award for Best Use of HPC in Physical Sciences'' \cite{HPC2018award}.

\section{Requirements and Design}
\label{sec:requirements}
The design of the Niagara supercomputer was a result of trying to satisfy a
number of requirements that were somewhat at odds.
The most important ones, were:

\subsubsection*{Requirement 1: A Large Parallel System}
The Niagara supercomputer has been designed to be a machine dedicated
to large parallel jobs with
intense computational and/or communication requirements.
\subsubsection*{Requirement 2: Satisfy Computing Needs of Existing Users}
In order to understand the needs of researchers, we ran a targeted survey asking
the ``biggest'' users
(i.e. the users that run the largest jobs and used a significant part of the overall resources;
for instance, we did not consider users running large serial jobs)
of our previous systems a set of questions to identify what type
of system they would benefit from most: a hybrid architecture with accelerators/GPUs,
a more homogeneous type of cluster, or other possibilities.
Most of the responses obtained reflected the need for a homogeneous
system with a high-performance interconnect
favoring MPI-type jobs.
\subsubsection*{Requirement 3: Smooth Transition from Previous Systems}
\begin{figure}
	\includegraphics[width=\columnwidth]{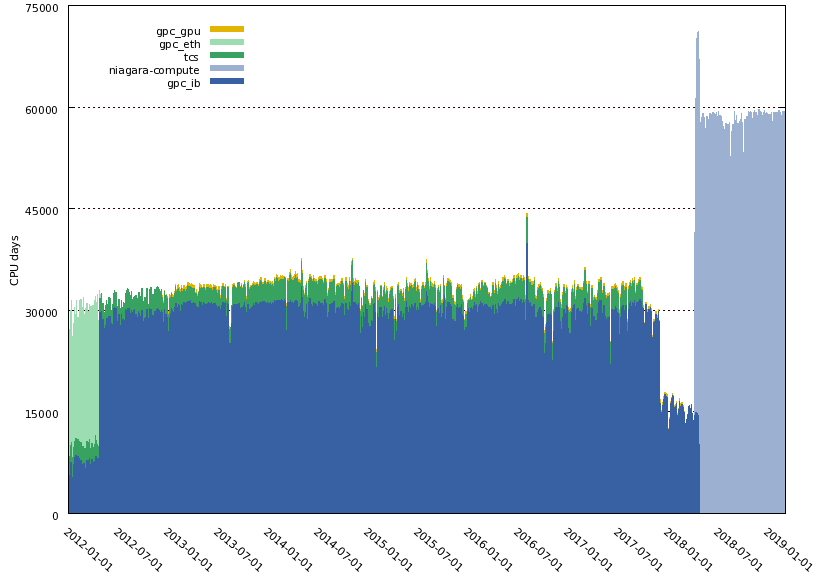}
	\caption{The utilization over time in CPU-days for the different SciNet's
		systems: GPC (30912 cores, 328 TFlops of peak performance), TCS (3264 cores, 60 TFlops of peak performance) and Niagara (60,000 cores, 4.6 PFlops of peak performance).}
	\label{fig:utilization}
\end{figure}

SciNet has been hosting several systems since 2008, that have been available to Canadian researchers across the country.
Among the most used ones were the GPC (a 30,912-core x86 cluster) and the
TCS (a 3264-core IBM Power 6 cluster)\cite{scinet-gpc}, of which the utilization has
been above 99\% (see Fig.~\ref{fig:utilization}).
The deployment of the Niagara supercomputer was designed to take into consideration the migration and transition
of users to the new system, minimizing disruption and interruption of their work.
\subsubsection*{Requirement 4: Avoid Issues Encountered on Previous Systems}
Using the experience of running high-performance clusters for more than a decade,
some measures were put in place in order to minimize potential flaws
in the design of the new system.
This affected the choice of
hardware, memory per core, interconnect,
file system capabilities, the directory structure
in user storage spaces, scheduling policies, etc.
For example, it is tempting to use features of
the scheduler to design and implement detailed policies regarding jobs and
priorities, but more constraints on the scheduling can make it hard to
get good utilization and makes it harder to debug the scheduling if it
does not do what was intended.
We also wanted to continue to offer an agile, optimized software stack, but previously lacked an automated
and well-documented process for installing software, which we intended
to improve upon.
\subsubsection*{Requirement 5: Compatibility and Coordination with Other Canadian National Systems}
The Niagara supercomputer is part of a national platform of
computational resources consisting, in addition to Niagara, of three
heterogeneous systems and several cloud systems,
available for Canadian researchers all over the country.
The new system was to be integrated with the national authentication
system, so there would no longer be a need for separate accounts for
each system, but it should also allow for `local' users that did not fall under
the national platform.
User support was expanded to allow specific staff members from
other consortia access to users' accounts when dealing with a support
ticket.  The national support ticketing system was augmented with a
Niagara `queue' and its capabilities were augmented to support the
team-based user support model of SciNet.
In addition to the specialized \textit{native} Niagara software stack,
the system should enable an almost transparent transition to users with a very similar
work-flow on Niagara as on the heterogeneous national systems,
by mounting the software stack available on those systems on Niagara (see Sec.~\ref{sec:software-stack}).

The design of Niagara emerged as a consequence of
contemplating and integrating the above mentioned requirements.
The technical details and hardware specifications of Niagara are presented and
described in the following section, while subsequent sections describe
the configurational solutions to the above requirements.
Of course, one should also mention that the additional budgetary constraint
of 20 million Canadian dollars (approx. 15 million USD).

\section{Hardware Selection}
 
\subsection{Design and Procurement}

\subsubsection{User Consultation Results}

As mentioned above, before starting the design of the new system, we consultated with 
our biggest users.  The results of these consultations made it clear
that a lot of those researchers preferred a homogeneous CPU cluster with a
fast interconnect rather that a hybrid or pure GPU.  However, we
did not want to impose this as a requirement in the RFP (``Request for
Proposal'') call.  To maximize the utility of the new cluster, we used
the results of the consultation to compile a series of representative benchmarks that were used as a baseline
to rank the different systems and configurations proposed by
interested vendors. 

\subsubsection{Benchmarks for a Large Parallel System}
The following application benchmarks make up the Large Parallel Benchmark (LPBM):
HPCG, Nek5000, WRF, NAMD, miniDFT, SPEC MPI 2007 and IOR. These
benchmarks were part of the requirements for potential vendors.
For each benchmark, we included the source code, run requirements and instructions for reporting the results.
In the following paragraphs, we provide details about all these benchmarks.

\begin{itemize}[leftmargin=*]
\item The High Performance Conjugate Gradient (HPCG) benchmark is designed to exercise
computational and data access patterns that closely match a broad set of important
scientific applications, and to give incentive to computer system designers to
invest in capabilities that will have impact on the collective performance of these applications.
	
\item Nek5000 is an open source, highly scalable and portable spectral element code designed to simulate a range of flow physics.
	
\item The Weather Research and Forecasting (WRF) Model is a next-generation mesoscale numerical weather prediction system designed for both atmospheric research and operational forecasting needs.
	
\item NAMD is a parallel molecular dynamics code designed for high-performance simulation of large biomolecular systems.
	
\item MiniDFT is a plane-wave density functional theory (DFT) mini-app for modeling materials. Given a set of atomic coordinates and pseudo-potentials, MiniDFT computes self-consistent solutions of the Kohn-Sham equations using either the LDA or PBE exchange-correlation functionals.
	
\item SPEC MPI 2007 focuses on performance of compute intensive applications using the Message-Passing Interface (MPI) library. 
	
\item IOR measures parallel file system I/O performance at both the POSIX and MPI-IO level.
This parallel program performs writes and reads to/from files under several sets of conditions and reports the resulting throughput rates.
				\end{itemize}

\subsubsection{Proposals Evaluation}
The LPBM provided a numerical measure of the scientific capacity
and capability of the proposed system.  It is related to the Scalable System
Improvement (SSI\footnote{SSI was built on the ``Sustained System Performance'' (SSP) Metric \cite{osti_861982}.}) metric \cite{8641549}
used in the NERSC\footnote{\url{https://www.nersc.gov/research-and-development/benchmarking-and-workload-characterization/ssi/}} and
 Crossroads\footnote{\url{https://www.lanl.gov/projects/crossroads/benchmarks-performance-analysis.php}} acquisition process in
that it normalizes the benchmarks against an existing system, in this case the
GPC, with the output being a single value of throughput and performance
improvement.
Vendors were required to run the benchmarks on a minimum of 100 nodes on their
proposed systems or prototypes and provide their LPBM score. The LPBM
score was combined with points for
technical merit, energy efficiency, implementation plan,
service levels and warranty, and vendor experience and
qualifications, into a single metric allowing the proposals to be
compared directly.    
 
We received 11 proposals from 6 vendors, a shortlist of the top 5 was selected,
and they were invited for an interview.
After the interview the final 5 vendors were allowed to make a final submission
and were scored. 
The top scoring vendor, in this case Lenovo, was selected to commence negotiations
which led to entering into a contract to provide the new system.
Final contract signing occurred in October 2017 with initial hardware delivery
commencing in December 2017, and full system acceptance in March 2018.

\subsection{Resulting Hardware Specifications}
The Niagara supercomputer is a large cluster of 1500 Lenovo SD530 servers each with
40 Intel Xeon Gold 6148 ``Skylake'' cores at 2.4 GHz (see Table~\ref{table:specs} for details).
The $Rmax$ peak performance, as measured by the HPL benchmark, of the cluster is 3.02
petaflops delivered with an $Rmax$ of 4.6 petaflops theoretical, which put it in the \#53
spot in the TOP500 list of June 2018.
Each node of the cluster has 187 GiB per node, i.e., well over 4 GiB/core
for user codes.

Niagara's operating system is Linux CentOS 7.4, which is a usual standard in the industry
and is also the supported OS by the vendors.

\begin{figure}
	\includegraphics[width=\columnwidth]{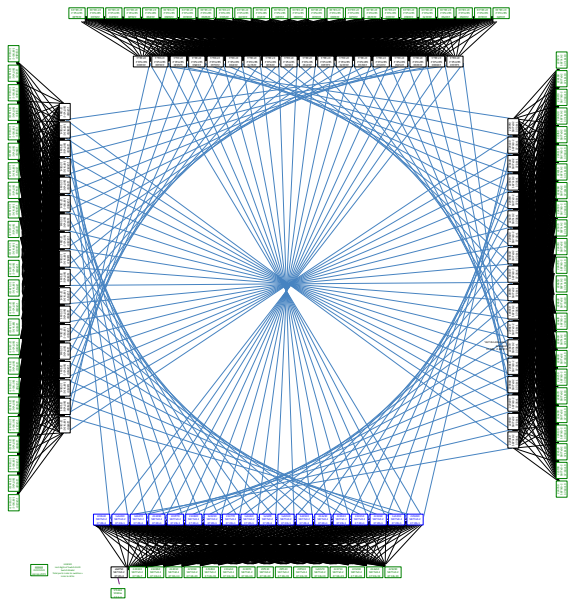}
	\caption{Connectivity diagram of Niagara's network interconnect,
	of Mellanox EDR InfiniBand in a ``Dragonfly+'' topology \cite{kim2010dragonfly,shipner2017}
	with 4 wings.}
	\label{fig:DragonFly-network}
\end{figure}

Designed for large parallel workloads, Niagara has a low-latency high-bandwidth
Mellanox EDR InfiniBand interconnect in a ``Dragonfly+'' topology \cite{kim2010dragonfly,shipner2017}
with 4 wings. A connectivity diagram is presented in Fig.~\ref{fig:DragonFly-network}.
Each wing of at most 432 nodes (i.e. 17280 cores) has one-to-one connections between leaf and core switches.
Network traffic between wings is handled by interconnections between core switches and with the
benefit of \textit{adaptive routing} \cite{bloch2013high}, which helps alleviate network congestion and balance load
across core interconnections, results in an effective blocking of 2:1 between nodes in
different wings.
Furthermore the Switch-IB 2 switches have on-hardware support for collective communication acceleration,
such as global reductions, which can significantly improve the overall interconnect performance in many HPC workloads.

The system has 12PB of raw storage, configured into approximately 9PB
of usable shared parallel file system, i.e.
IBM Spectrum Scale --commonly referred to as IBM General Parallel File System (GPFS)--,
that is mounted on all nodes.
Additionally a 256TB Excelero burst buffer (NVMe fabric, up to 160 GB/s) is available for fast I/O operations.

Finally, a set of 28 management nodes for logins (8), data movement
(2), GPFS (4),
subnet managers (2), xcat managers (2), and service nodes (8)
complete the setup. These management nodes have 20 Skylake cores of
the same architecture as the compute nodes, and 96 GB of RAM.

\begin{small}
\begin{table}
\begin{tabular}{p{0.95\columnwidth}}
\toprule
    Compute Nodes - Lenovo SD530 with 2$\times$ Intel Xeon Gold 6148 ``Skylake'' (20 cores, 2.4GHz) \& 192 GB (1500 nodes, 60,000 cores)
\\
\hline
    Management Nodes - Lenovo SR630 with 2$\times$ Intel Xeon Gold 5115 ``Skylake'' (10 cores, 2.4GHz) \& 96GB (28 nodes):
%\\
    2$\times$ Datamovers,
%\\
    4$\times$ GPFS,
%\\
    8$\times$ Login,
%\\
    2$\times$ Scheduler,
%\\
    2$\times$ xCAT Manager,
%\\
    2$\times$ Subnet Manager,
%\\
    8$\times$ xCAT Service
\\
\hline
Mellanox 4-wing Dragonfly+ EDR (ConnectX-5 \& Switch IB2) Infiniband
\\
Core: 72$\times$ Mellanox SB7890 EDR IB Switch (36 ports)
\\
Leaf: 84$\times$ Mellanox SB7890 EDR IB Switch (36 ports)
\\
\hline
Lenovo DSS-G260 (504$\times$10TB) 4PB usable $\times$3 = 12PB /scratch \& /project (GPFS)

Burst Buffer - Excelero NVMe over fabric (IB), 256TB (10$\times$ SR650 nodes with 8$\times$6.4TB Intel P4600) (GPFS)
\\
\bottomrule
\end{tabular}
\caption{Technical specifications and configuration of the Niagara
  supercomputer (excluding the test development system).}
\label{table:specs}
\end{table}
\end{small}

\subsection{Deployment}
In order to test the configuration, software stack and
custom-developed build system, we created a virtual cluster, and then
a small version of the proposed one, the Test and Development System
(TDS).  The TDS consists of 4$\times$ Lenovo SD530 Nodes (40 core,
192GB RAM).  This is the exact same hardware configuration as the 
for 1500 compute nodes of Niagara was going to have.  Here we were able to test and
develop different configurations before implementing it in the actual
cluster.
Even with Niagara in production, this TDS remains a
crucial sandbox when rolling out OS updates, major
changes in the software stack, or scheduler modifications.

\begin{figure}
        \includegraphics[width=\columnwidth]{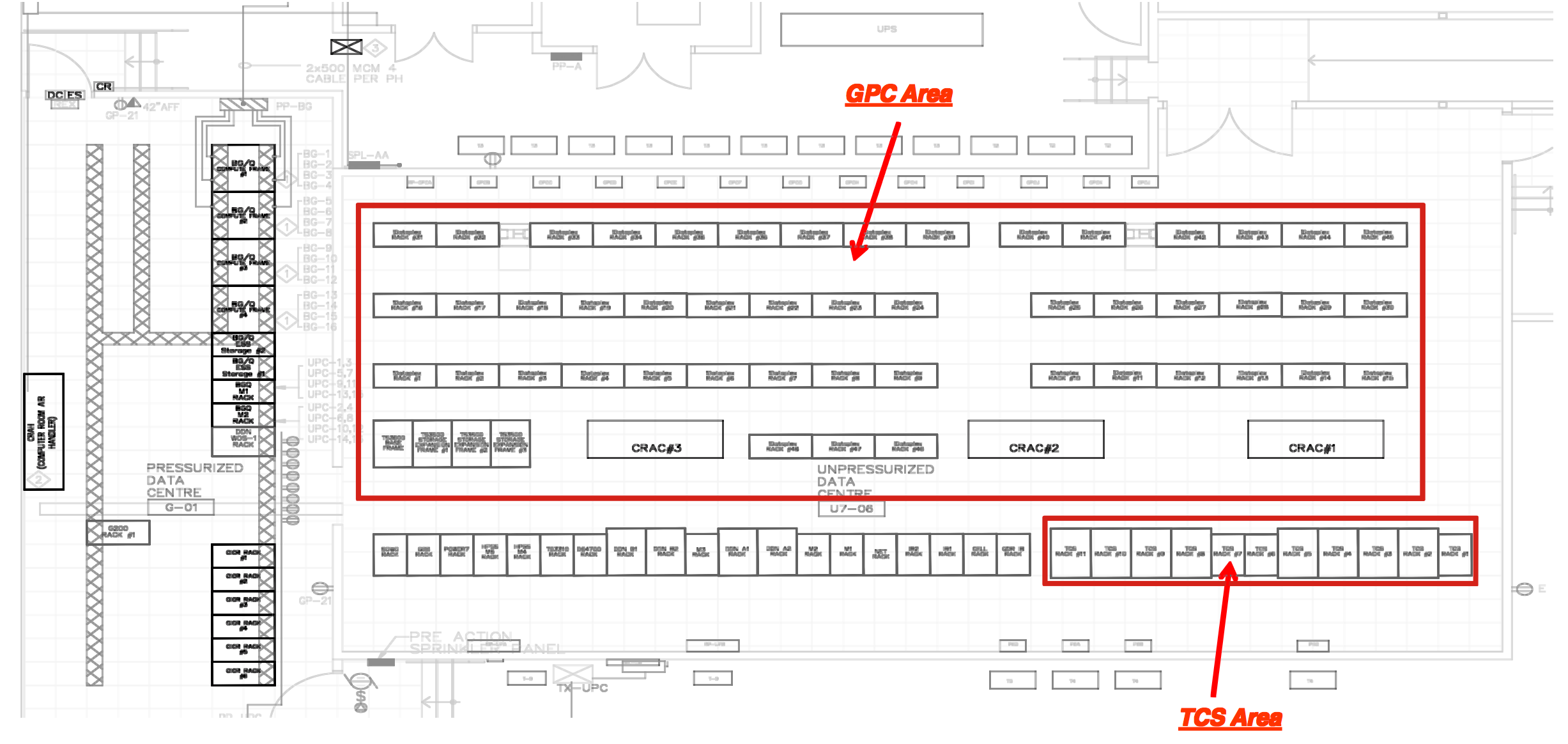}
        \includegraphics[width=\columnwidth]{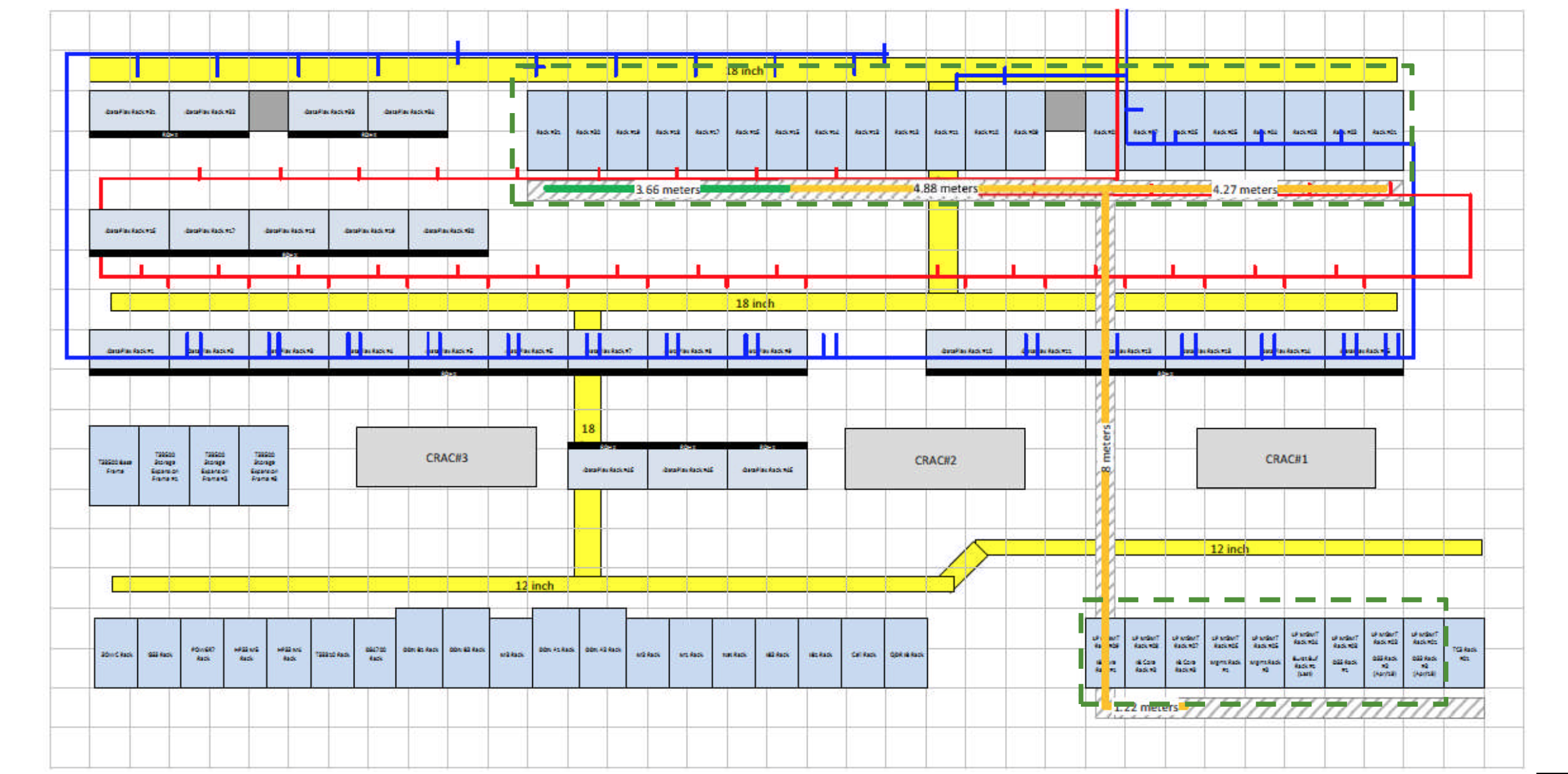}
        \caption{Floorplan of SciNet's datacenter --
                top: layout of the GPC and TCS clusters, previous to Niagara installation;
                bottom: the Niagara system took the space of the TCS (for management nodes) and
                half of the GPC (for compute nodes). Additionally modifications where needed for
                electric supply and water cooling pipes.
                A time-lapse movie of the removal and installation process can be seen at
                \url{https://www.scinet.utoronto.ca/launch-of-niagara/}.}
        \label{fig:floorplan}
\end{figure}

Before the actual deployment of Niagara, we needed to plan and execute
the decommissioning of our two previous clusters, the TCS and the
GPC, in order to make room for the new equipment, and the required
modification of the datacenter to accommodate  the new cluster.
Fig.~\ref{fig:floorplan} shows the floorplan of SciNet's datacenter, mapping
the space where the old TCS and GPC systems were located and the space taken
for Niagara.
Although most of the datacenter's existing infrastructure (power supply,
water cooling, chiller, etc.) could be reused, a number of
invasive renovations were required, such as 
removing not only electrical subfloor
cables but also water pipes for cooling the previous systems.
The whole renovation process took about a month, starting in late October, and by the end
of November Niagara's equipment started to arrive at the data center.
By early December most of the equipment had arrived at the site and the actual
installation began.

We first removed all of the TCS
and half of the GPC nodes.  In this way, users could still use the remaining
GPC while we transitioned to the new cluster. This was made possible by the
higher computational density of Niagara, which provides 
roughly ten times the computational power in only half of the physical space.
Niagara's installation began with the management, storage and core network nodes,
which were placed in the space of the old TCS (see
Fig.~\ref{fig:floorplan}). After half of the GPC was removed, the Niagara compute nodes were put in place. The system was ready for testing in early January of 2019, first by the vendor, then by SciNet and a few early users (see Sec.~\ref{sec:science}).  

Most of the rest of the GPC was removed afterwards, but a small part,
42 nodes with 16 cores each, was repurposed as a teaching cluster. It
uses the same configuration as Niagara, and mounts the same file
systems.  There has been a remarkable demand for this HPC resource for teaching; in its first academic year, the teaching cluster has already been used by five courses at the University of Toronto.

\section{Software Stack}
\label{sec:software-stack}
To give access to a large variety of distinct software packages to our users, we deployed a \textit{module-based} software stack (a common practice on supercomputing facilities).
Modules are configuration files that contain instructions for modifying the users' software environment.
This modular architecture allows multiple versions of the same application to be installed without conflicts.

Our previous system's software stack did not have an automated or
systematic software installation procedure, which in the course of
time caused some maintenance headaches. After reviewing available
automated software installation systems, we decided to develop our own tool for
building and maintaining a software stack of modules systematically organized
and documented for ease of rebuilding, reinstallation and upgrading
software.	
This agile HPC cluster software build system is called
 \textit{Cooperage} (\url{https://gitlab.com/scinet-hpc/cooperage}). It ties into the software modules system
Lmod \cite{McLay:2011:BPD:2063348.2063360} developed at TACC (\url{https://www.tacc.utexas.edu}).
Cooperage is akin to alternatives, such as Easybuild and Spack, but is
simpler in its approach and usage, less generic (albeit with fewer features), more tailored to Niagara.
This allows us to quickly deploy a cutting edge software stack and makes
easy to port the software installation to compatible clusters.

In order to satisfy some of the requirements mentioned in Sec.~\ref{sec:requirements},
Niagara in fact offers two software stacks:
\begin{enumerate}
    \item A ``native'' Niagara software stack tuned and compiled for Niagara's specific architecture,
which is built using Cooperage.
    \item The same software stack available on other general purpose systems in the country.
\end{enumerate}

Users can switch between these two stacks using the same Lmod modules system.
Deploying the ``general purpose'' stack was not trivial, as it is
distributed using CVMFS \cite{Buncic_2010,Blomer:2011:CDS:2110217.2110225},
whose standard mode of operation requires node-local storage, which
Niagara does not have.  The stack was deployed using CVMFS's
`alien cache' technology \cite{5560054} that makes it possible to run CVMFS on
systems without local disks on each node and also leverages the performance
available from the large GPFS file system.
This setup means that users can experience slight lags in updates to the stack, depending
on the frequency of pulls from the main repository, but in practice
this only  has
a small impact.

The Lmod installation is used by both
stacks, but with different environment variables pointing at the appropriate
configuration and module paths.
\footnote{
As a general rule, we advice users not to load modules in their .bashrc as this
may cause side effects when the .bashrc file is loaded in other circumstances, or as the modules required change in the course of a project.
Instead, we recommend to load modules on the command line when needed, or by sourcing a
separate script, and to do the same in job submission
scripts.}

\section{Job Scheduler}
\label{sec:scheduler}

We use SLURM (Simple Linux Utility for Resource Manager)~\cite{10.1007/10968987_3}
as our job scheduler and resource manager. Its basic task is to track the 
available resources at current and future times and allocate them to jobs
waiting in a queue according to a priority assigned to them.

SLURM is extremely flexible on the way it can be configured.
However, in our experience, the fewer constraints one puts on a
scheduler, the better it can do its job of delivering high and fair
utilization of the cluster. Therefore, we started with the simplest
possible scheduler configuration and only added complexity or new
constraints when usage data showed significant deviations from the
expected fair usage. 

We deployed Niagara with 
a minimum number of \textit{partitions} (queues in SLURM nomenclature)
to satisfy the users' needs. The vast majority of users do not  
have to worry about partitions at all. If their scripts do not choose 
a partition the default partition, called \textit{compute}, is assigned to the job. 
Each job must request at least one node and cannot request more than
1000 nodes (larger computations are scheduled manually). Nodes are allocated exclusively to the job for 
a minimum of 15 minutes and a maximum of 24 hours. The
latter restriction ensures a reasonable turn-over of jobs and makes
large jobs feasible without lowering utilization. 
Of the 1500 nodes, 1495 are available to jobs submitted to this 
partition. The remaining 5 nodes are exclusively used by the \textit{debug}
partition, which can use any of the 1500 compute nodes. 
We impose a limit of 1 hour and a maximum of 1 job 
per user for jobs in the debug partition. Finally, we have 4 partitions matching 
each of the 4 wings in our Dragonfly+ network topology (named \textit{dragonfly[1-4]})
that exist to satisfy the needs of some groups that would like to run their codes 
on one of the wings to take full advantage of 1:1 network blocking.
In the end, however, only the compute
and debug partitions are used by the vast majority of jobs.

In addition, there are
\textit{archive partitions} to interact with the HPSS
archiving system and allow users to transfer their data to the tape
system. Their configuration differs in some aspects
from the compute partitions. We allow several jobs to share the 
same archive node interacting with the HPSS system, for example.
In one partition, users are allowed up to 75 jobs with a maximum
of 1 hour.  In another partition, there is a limit of 3 days and up
to 5 simultaneous jobs, for larger data recalls. Finally, we have a 
partition with a limit of 1 hour and up to 48 simultaneous jobs,
intended for interactive usage of the HPSS system.

We use the SLURM multifactor priority plugin to balance the various
factors used in the priority computation, such as the job age and size,
the partition it was submitted to, the job's quality of service, and the user's fair-share of the system. The main factor in this
computation is the fair-share. Once a year in the fall, there is a call
for proposals for resource allocations on the national systems, which is coordinated by Compute Canada. Those proposals are evaluated both
technically and scientifically. In the fair-share computation, the scheduler takes into account 
the group's usage in the previous 7 days and adjust the priority 
factor accordingly to that group's target share of the system.
About 6\% of the system is not allocated, and can be used by
so-called default
users, those that either did not apply or were not awarded a resource allocation.
If groups with an allocation underutilize their share, others
can use those compute cycles.  
The job age in the queue has an effect on the priority but it reaches 
a maximum contribution after 14 days on the queue. The scheduler favors 
large jobs but we reduced its weight in the priority computation to avoid
large jobs from opportunistic users trumping smaller jobs from allocated users
with a larger fair-share of the system. We set the partition weight twice as 
large as the fair-share one,  so  that
interactive sessions will start promptly. Finally, we use the quality-of-service factor,
which we set as high as the fair-share one, to take care of special
cases, such as to bump some user's priority temporarily, to put a limit
on the number of nodes a default user can ask for, or to limit the number of jobs
someone can submit to the scheduler.

SLURM is shipped with a job submission plugin, which is very useful to parse 
job submission scripts and act as a filter to catch 
common mistakes. However due to some shortcomings at the time of Niagara's
deployment, we ended up writing our own job submission wrapper python script.
This script deals with local peculiarities of our cluster configuration 
and either alerts the user of a possible unintended consequence or aborts 
the submission altogether with a meaningful error message. For example,
the home file system is read-only from the compute nodes. If a user tries
to submit a job from the home file system this user will get a warning. 

There are a few characteristics of our cluster configuration which affects
how jobs execute. The first one worth mentioning regards hyperthreading.
It is a technology that leverages more of the physical hardware by considering
twice as many logical cores than existing physical cores. Both the operating system
and SLURM see twice as many cores when scheduling jobs. It is a technology that
depending on the application, for example, most of cpu-bound codes reaches a 5-10\% speedup.
Since we schedule jobs by node, it is fairly easy for the user to use hyperthreading.
It just requires a change in the number of tasks from 40 to 80 per node and
the extra logical cores come as a bonus.
The second aspect of our configuration worth mentioning is that compute nodes
have no internet access. Therefore data stored outside our intranet needs to 
be downloaded before submitting a job. We impose this limitation to avoid 
waste of computational resources while outside data is downloaded.
Similarly software requiring license authentication via outside servers might 
need to follow a special protocol to overcome this limitation such as 
establishing a ssh tunnel in order to gain access to the outside world. 
Finally, we should mention that our SLURM version at the time of deployment, 
17.11.5, didn't have the native X11 support working.
We backported most of SchedMD's recent patches, adapting them to our system
(for example, not allowing the .Xauthority file to be written on home file system from the compute nodes.)
As a result we have the Slurm native X11 support working satisfactorily.

\section{File Systems}
\label{sec:FS}
With the exception of our tape system, the file systems on SciNet use GPFS \cite{gupta2011gpfs},
a high-performance file system which provides rapid reads and writes to large
datasets in parallel from many nodes.

Niagara provides several different file systems,
summarized in Table~\ref{table:FS}.

\begin{small}
\begin{table}[!h]
\begin{minipage}{\columnwidth}
\begin{tabular}{|p{0.115\columnwidth}||p{0.125\columnwidth}|p{0.075\columnwidth}|p{0.135\columnwidth}|p{0.085\columnwidth}|p{0.085\columnwidth}|p{0.115\columnwidth}|}
    \toprule
    File System/Set  &       Quota           &  Block Size   &   Purging Policy & backed up & on login nodes & on compute nodes  \\
    \hline\hline
    \rowcolor{gray!50}
    /home          &   100\;GB per\;user     &  1 MB         &                                  &    yes    &    yes &   read-only           \\
    \hline
    \rowcolor{gray!50}
    /scratch       &   25\;TB per\;user
                                         	& 16 MB         &
    files not accessed in 60 days           &   no  &   yes &   yes \\
    \hline
    /project       &   by group allocation & 16 MB         &                   &   yes & yes   & yes   \\
    \hline
    /bb            &       10\;TB per\;user  & 1 MB          &   very short      &    no &   yes &   yes  \\
    \hline
    HPSS       &   by group allocation &   &   &   dual-copy & no  &   no \\
    \bottomrule
\end{tabular}
\caption{File System and Storage Policies.
{\footnotesize Shaded rows represent FS with default access to any user; while non-shaded ones are per request.}
}
\label{table:FS}
\end{minipage}
\end{table}
\end{small}

/home is intended primarily for individual user files, common software
or small datasets used by others in the same group. 
/scratch is a larger storage space to be used primarily for temporary or transient files, for all the results of computations and simulations, or any material that can be easily recreated or reacquired. 

/project is intended for common group software, large static datasets, or any material very costly to be reacquired or re-generated by the group. In contrast to /scratch, /project is backed-up. Material on /project is expected to remain relatively immutable.
/bb, the \textit{Burst Buffer}, is a very fast, very high performance alternative to scratch, made of solid-state drives (SSD) \cite{excelero}.

HPSS is a nearline storage pool, for offloading semi-active material from any of the above file systems.  This was an existing storage solution that was migrated to Niagara.

\subsection{Burst Buffer}
One of the unique features of the Niagara cluster is the so-called
\textit{burst buffer}.
Niagara's burst buffer is a fast, high performance shared file system, made of
solid-state drives (SSD). While the overall bandwidth of the burst buffer is
comparable to the one of the scratch file system, the true strength of the
burst buffer lies in dealing with high I/O operations per seconds (IOPS). The
ideal use-cases are therefore jobs which involve a lot of IOPS, too many for
the /scratch file system, such as certain bio-informatics workflows and quantum
chemistry calculations, and codes that have large restart checkpoint files to
be saved/read between jobs.

The setup of the Burst Buffer of the Niagara cluster is evolving as we come to
better understand how to use best this resource. The current setup is described
below.

\subsubsection{Short-term persistent burst buffer space}
To get access to space on the burst buffer that is persistent between jobs, a
user must first request space on it.  Users with short-term persistent burst
buffer access will have a directory created on that resource.
The location is accessible using the \texttt{\$BBUFFER} environment variable.
Unlike ramdisk or job-specific burst buffer space (explained below), the files
will remain on the user's persistence burst buffer space between jobs. This makes
persistent burst buffer ideal for codes that have large restart checkpoint
files to be saved between jobs.

The persistence of files on this burst buffer space is very limited, so users
should still endeavour to clean up after each job, by staging out final files
to /scratch and removing temporary files.

\subsubsection{Per-job temporary burst buffer space}
For every job on Niagara, the scheduler creates a temporary directory on the
burst buffer called \texttt{\$BB\_JOB\_DIR}. The \texttt{\$BB\_JOB\_DIR}
directory will be empty when a job starts and its content gets deleted
after the job has finished. This directory is accessible from all the nodes of
a job.

\texttt{\$BB\_JOB\_DIR} is intended as a place for applications that generate
many small temporary files or that create files that are accessed very
frequently (i.e., high IOPS applications), but that do not fit in ramdisk.

It should be emphasized that if the temporary files do fit in ramdisk, then
that is generally a better location for them as both the bandwidth and IOPS of
ramdisk far exceeds that of the burst buffer. To use ramdisk, the user can either
directly access \texttt{/dev/shm} or use the environment variable
\texttt{\$SLURM\_TMPDIR}.

\subsection{File System Layout}
Another critical point in terms of the file system performance is the design of 
the layout for the different file system spaces themselves.
Having already resolved performance issues on the previous systems with a flat organization of
thousands of user home and scratch directories (eg. {\small
  /home/\$USER} and {\small {/scratch/\$USER)),
by switching to a hierarchical organization, on
Niagara the storage layout was also organized in a hierarchical
fashion, by group name
and by first letter of the group, so that a user's home directory
lives at {\tt \$HOME = /home/\$\{G\}/\$GROUP/\$USER} and their scratch
directory at {\tt \$SCRATCH = /scratch/\$\{G\}/\$GROUP/\$USER}.
Such a layout allows to aggregate users under a particular principal investigator.
 Because of this, it also helps in the process of setting an uniform set of permissions within the group.
 Moreover, and of critical importance on the performance side, the natural categorization
 improves substantially the parallel file system responsiveness as a whole in contrast with
 a flat hierarchy where conflicting directory blocking could easily occur.

\section{Monitoring Tools}
\label{sec:monitoring}

We use standard monitoring tools such as Nagios \cite{Josephsen:2007:BMI:1208311},
Ganglia \cite{massie2004ganglia} and Munin \cite{munin} to monitor the data center and cluster status.

In addition to that, we have developed system specific tools for users to keep track of
the system and resources utilization.
We offer them via a website interface, \url{my.scinet.utoronto.ca} and also as
command-line tools.

\subsection{my.scinet Portal}
\texttt{\url{my.scinet.utoronto.ca}} is a web portal that allows users to access real-time information
about Niagara's utilization and in particular the user's jobs.
The website offers a quick overview of Niagara (Fig.~\ref{fig:myscinet}), displaying interactive
visualizations with an overall status of the Niagara supercomputer, by showing how much computation
is being done and how much file storage is being utilized.
There is also information about job statuses, expected turnover times for jobs and compute nodes,
as well as job distributions as a function of time.
Similarly, information about the resources (memory, CPU and overall load) on the login nodes
is available at the bottom of the page.
In addition to that, the users are able to log into the portal with their system credentials,
gaining access to information about the user's resource allocation, storage utilization and
past jobs. In particular, job performance in terms of core utilization, communication, IOPs, FLOPS, etc.
The actual website is an elixir/phoenix app, with
 a Postgres/TimescaleDB backend.
The source is available at
	\url{https://gitlab.com/scinet-hpc/myscinet}.

The \texttt{\url{my.scinet.utoronto.ca}} web portal not only provides user's
access to their job's performance data, it has also been a valuable
tool for analysts to identify less than
optimal usage of the system.

\begin{figure}
	\includegraphics[width=.495\columnwidth]{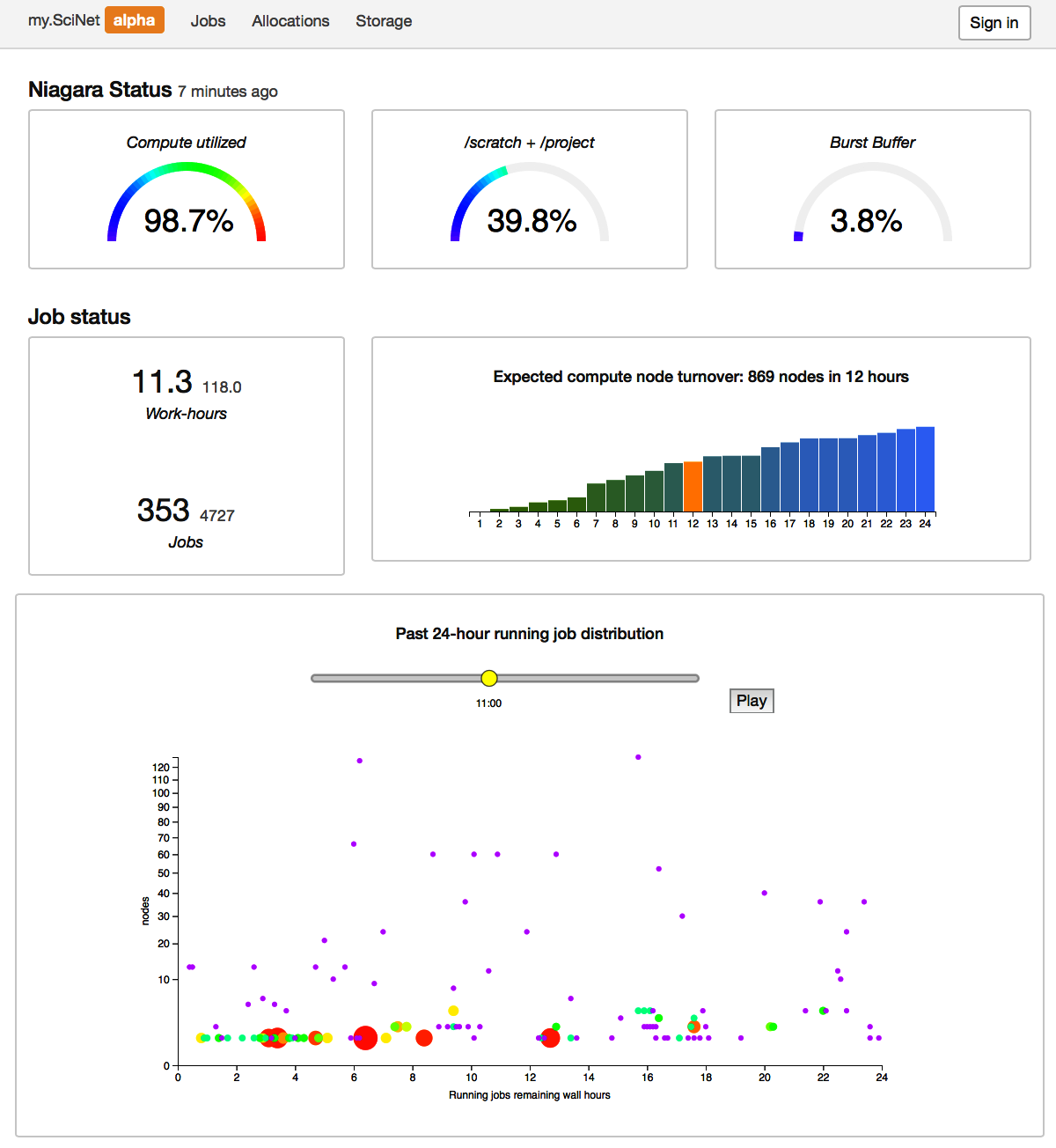}
	\includegraphics[width=.495\columnwidth]{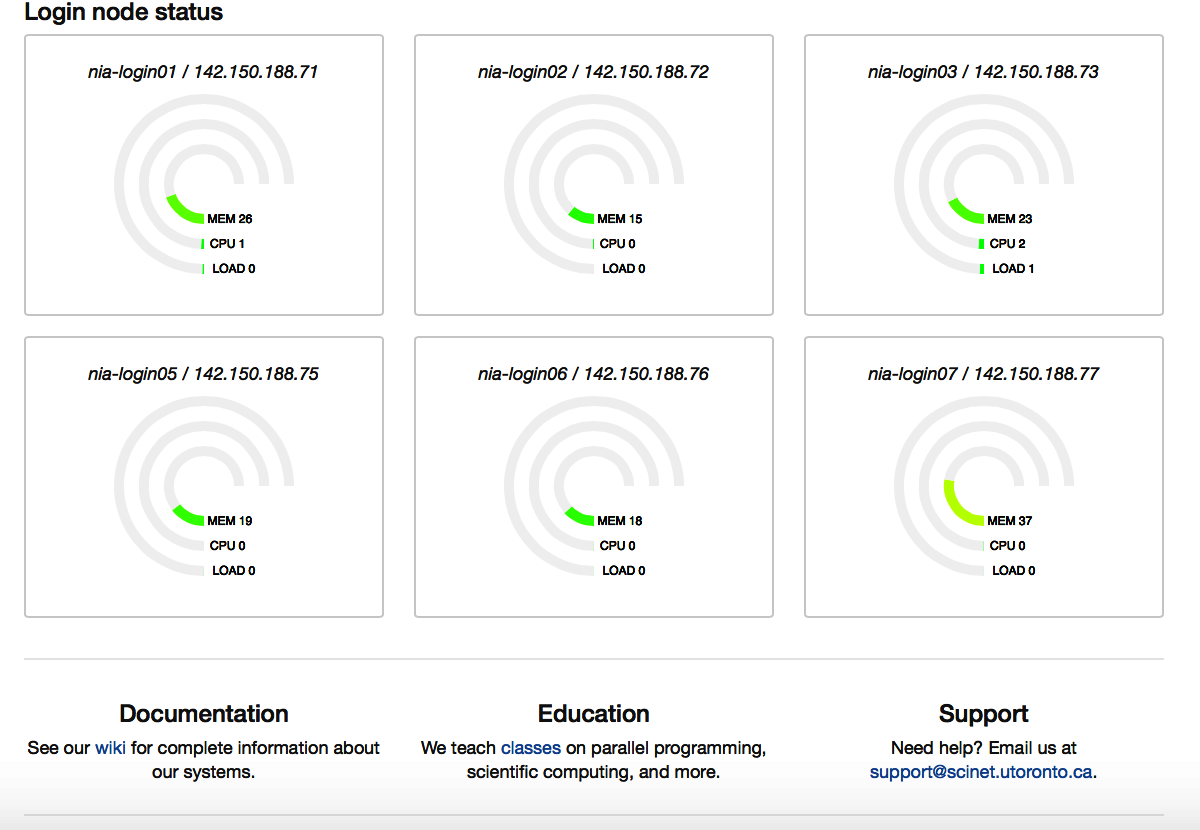}
	\caption{Real-time information presented in the \url{my.scinet.utoronto.ca} portal,
	offering a graphical overview of the Niagara system status and utilization.}
	\label{fig:myscinet}
\end{figure}

\subsection{Command-line Tools}

SLURM lets you look at the queue of jobs with
the squeue command. In addition, we wrote a utility, called qsum, that allows us to look
at the queue with an emphasis on the user instead of jobs. Qsum summarizes
the queue by user and lets us know at a glance how many cores and jobs a particular
user is running or pending and how long on average those jobs are going to take. 
It also helps us to identify which fraction of the system is being used 
by our default users at that moment in time. It is a way of evaluating how fair 
the system is being allocated with respect to the default users.  

There are a few other custom commands that are accessible at the
command-line level for users to query the 
following aspects:
\begin{small}
\begin{description}
        \item[\texttt{jobperf}]
		query the performance of a running job
        \item[\texttt{nodeperf}]
		query the performance of an specific node
        \item[\texttt{diskUsage}]
		query the user's filesystems utilization
        \item[\texttt{scinet niagara priority}]
		provides information of the user's allocation utilization and scheduling priority
\end{description}
\end{small}

Most of these tools and others, such as wrappers around the sbatch scheduler
handler -- for instance, adding enhanced error and warning messages,
or an explicit command, \texttt{debugjob}, for requesting an interactive job,
can be found in our repository,
	\url{https://gitlab.com/scinet-hpc/niagara-utils}.

\section{Early Science Cases}
\label{sec:science}
In order to test and fine-tune the system, during a short period of time
immediately after its deployment by March 2018, a select number of
scientists were given the opportunity to perform what we called ``heroic'' calculations.
Utilizing nearly the entire system
these large scale calculations were essential to test, to tune and to get Niagara ready
for use as Canada's fastest national academic supercomputer.
In addition to the scientists performing the simulations, the process also
involved personnel from our analyst team helping to have the basic tools available on the
system, as well as, troubleshooting any potential issues that may have arisen
as a consequence of having the system almost in its bare-bones.
Personnel from the vendors was also involved, helping in particular to
debug and improve different aspects and configurations in the system.

Four different science cases were pursued in the early stage of the cluster:
 a global ocean's circulation model with super-high spatial and temporal resolution,
to resolve the nature of the turbulence near the ocean's floor \cite{heroicRun,HPC2018award};
 a stellar hydrodynamics simulation of core convection in a 25 solar mass star on a
1536-cubed grid followed over 57.7 days of star time \cite{2018arXiv181013416W,HerwigRD};
 a full numerical relativity simulation trying to trigger the onset of magnetorotational
instability during the merger of a binary neutron star system;
and a convergence study of global, general relativistic magnetohydrodynamic
simulations of non-radiative, magnetized disk modeling black hole accretion flows.
Although different in their respective foundations, algorithms and numerical implementations
and the actual science that these simulations were tackling, they share the common theme
of being truly demanding and intense computational problems.
Not only did these represent the cutting-edge in their respective scientific
fields, they stressed the performance of the system to its limits
(in particular the first two cases utilized almost the full system, i.e. 1200 nodes
and 1152 nodes respectively).
By running these simulations we were able to test the network connectivity, the 
IO responsiveness of the filesystem, fine-tune different configuration parameters
and find problems during the installation that otherwise we wouldn't have been able to detect.

\section{Transition and Migration}
Since the systems that Niagara replaced were highly utilized, it was
important to have as smooth a transition to the new system as
possible. 

Several weeks before the final shutdown of the GPC system, we began the process
of copying users' files and migrating accounts from the old GPC and TCS systems
to Niagara.
In particular, transferring files from the users took several iterations of
rsync-ing their contents followed by a short period of few days when users were
not allowed to log in the systems for a final pass and integrity check
of the information transferred.
At the same time, accounts from the previous systems were migrated into a new
LDAP configuration.

From a total of 2627 accounts present on the GPC at the moment of
decommissioning it, a total of 2216 accounts were migrated to Niagara.
So far we have a total number of 3104 accounts created on Niagara, from
which 2537 are currently enabled.
From these active accounts, 1690 were active on the GPC system,
which means that more than 1/3 ($\approx 35\%$) of the current accounts on
Niagara are new accounts.

\section{Conclusion}

In this paper we have described the procurement, installation
and configuration of the Niagara supercomputer.
We started by providing details in our design process and discussed the different
elements considered in order to define what type of cluster we wanted Niagara
to be.
Key elements in its design where the homogeneity across the cluster and
dominant high-speed interconnect given by a state-of-the-art DragonFly+
topology.
Other elements that make Niagara different and unique are: its high-speed optimized
high-IOPS burst buffer file system, the approach to its software stack and own
cluster build system, Cooperage.
We have presented our customized system of tools offered to users, and make them
available to the community via a public repository.
Niagara has been operating near maximum capacity from the
moment it was taken into general production (see Fig.\ref{fig:NIA-slurm-jobs}).
This plus the fact that, in a little over a year of operations we have
already passed the threshold of 1.2 million submitted jobs, represents a clear
indication of the need among Canadian researchers for a system capable of running
large parallel jobs.

\begin{figure}
        \includegraphics[width=\columnwidth]{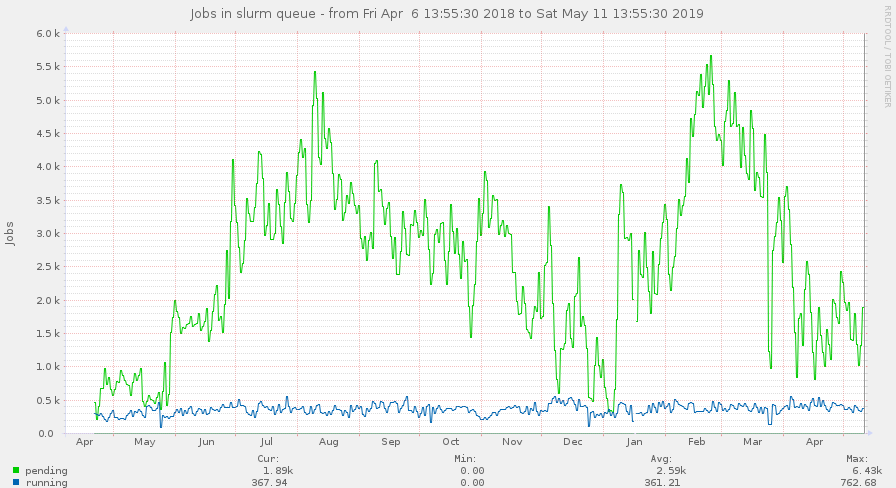}
        \caption{Jobs submitted to Niagara in the last year of operations, as reported by Munin \cite{munin}.}
        \label{fig:NIA-slurm-jobs}
\end{figure}

\begin{acks}
The Niagara supercomputer is jointly funded by the Canada Foundation for Innovation, the Government of Ontario, and the University of Toronto.
We would like to thank the support and help during the deployment process of Lenovo, Mellanox, and Excelero.
We would also like to acknowledge the researchers working in the early stages of development,
who helped to debug the cluster
 by attempting the largest possible
runs across the whole machine -- i.e.\ Prof. W.R.\ Peltier's group, Prof. F.\ Herwig's group, and Profs. L.\ Lehner and S.\ Liebling. 
\end{acks}

\bibliographystyle{ACM-Reference-Format}
\bibliography{refs}

\appendix
\input{onlineResources}

\end{document}

%% file: ACMcopyright.tex
\copyrightyear{2019}
\acmYear{2019}
\setcopyright{acmlicensed}
\acmConference[PEARC '19]{Practice and Experience in Advanced Research Computing}{July 28-August 1, 2019}{Chicago, IL, USA}
\acmBooktitle{Practice and Experience in Advanced Research Computing (PEARC '19), July 28-August 1, 2019, Chicago, IL, USA}
\acmPrice{15.00}
\acmDOI{10.1145/3332186.3332195}
\acmISBN{978-1-4503-7227-5/19/07}

%% file: authors1.tex
\author{Marcelo~Ponce,	%, %}  %\email{mponce@scinet.utoronto.ca, %}
Ramses~van~Zon,	%, %} %\email{rzon@scinet.utoronto.ca, %}
Scott~Northrup,	%, %}
Daniel~Gruner,	%, %}	%\email{dgruner@scinet.utoronto.ca, %}
Joseph~Chen, %}
Fatih~Ertinaz, %}
Alexey~Fedoseev, %}
Leslie~Groer, %}
Fei~Mao, %}
Bruno~C.~Mundim, %}
Mike~Nolta, %}
Jaime~Pinto, %}
Marco~Saldarriaga, %}
Vladimir Slavnic, %}
Erik~Spence, %}
Ching-Hsing~Yu, %}
W.~Richard~Peltier}

\email{support@scinet.utoronto.ca}

\affiliation{%
  \institution{SciNet HPC Consortium, University of Toronto}
  \streetaddress{661 University Ave, suite 1140}
  \city{Toronto}
  \state{Ontario}
  \postcode{M5G 1M1}
  \country{Canada}
}

%% file: onlineResources.tex
\section{SciNet/Niagara Online Resources}

\fbox{
\begin{minipage}{0.95\columnwidth}
%\begin{description}
%	\item
	Main website:	\quad
		\url{https://www.scinet.utoronto.ca}

%	\item
	Education and Training:	\quad
		\url{https://courses.scinet.utoronto.ca}

%	\item
	System status, and technical documentation:

	\quad	\url{https://docs.scinet.utoronto.ca}

%	\item
	System statistics:	\quad
		\url{https://my.scinet.utoronto.ca}

%	\item
	Niagara deployment timeline:

	\quad	\url{https://www.scinethpc.ca/road-to-niagara/}

%	\item
	Niagara story:

	\quad	\url{https://www.scinet.utoronto.ca/launch-of-niagara/}

%	\item
	SciNet HPC tools repository:	\quad
		\url{https://gitlab.com/scinet-hpc}
		%|
		%\url{ https://gitrepos.scinet.utoronto.ca/public/}
%\end{description}
\end{minipage}
}